\documentclass[12pt]{article}
\usepackage{amsmath}
\usepackage{graphicx,psfrag,epsf}
\usepackage{enumerate}

\usepackage{natbib}
\usepackage{amsfonts}
\newcommand{\B}{{\cal B}}
\newcommand{\R}{\mathbb{R}}
\newcommand{\conv}{\mbox{conv}}

\newcommand{\blind}{0}

\begin{document}

\def\spacingset#1{\renewcommand{\baselinestretch}%
{#1}\small\normalsize} \spacingset{1}


\if0\blind
{
  \title{\bf Improving the local scoring algorithm using gradient sampling}
  \author{M.-O. Boldi\thanks{Corresponding author (E-mail: {\it marc-olivier.boldi@unil.ch}, tel: +41223798837)}\hspace{.2cm}\\
    Faculty of Business and Economics, University of Lausanne\\
    and \\
    V. Chavez-Demoulin \\
    Faculty of Business and Economics, University of Lausanne}
  \maketitle
} \fi

\if1\blind
{
  \bigskip
  \bigskip
  \bigskip
  \begin{center}
    {\LARGE\bf Improving the local scoring algorithm using gradient sampling}
\end{center}
  \medskip
} \fi

\bigskip
\begin{abstract}
We adapt the gradient sampling algorithm to the local scoring algorithm to solve complex estimation problems based on an optimization of an objective function. This overcomes non-differentiability  and non-smoothness of the objective function. The new algorithm estimates the Clarke generalized subgradient used in the local scoring, thus reducing numerical instabilities. The method is applied to quantile regression and to the peaks-over-threshold method, as two examples. Real applications are provided for a retail store and temperature data analysis.
\end{abstract}

\noindent%

\spacingset{1.45}
\section{Introduction}
\label{sec:intro}
Among the statistical applications based on the optimization of an objective function (e.g., negative log-likelihood, risk function, etc.), more and more are based on a function that can be non-differentiable and/or non-convex. In addition, the parameter estimates are often constrained to have some predefined properties (e.g., being linear, additive in covariates, smooth, etc.). This paper introduces the use of the gradient sampling algorithm in such cases. 

In essence, the gradient sampling algorithm, introduced by \cite{BuLeOv2002}, is a descent algorithm where the gradient direction is replaced by a stochastic approximation of the Clarke subdifferential. As a result, the descent algorithm is stabilized and can be used for non-convex and/or non-differentiable objective functions. To illustrate, in this paper, we propose the use of the gradient sampling approximation for the local scoring algorithm of \cite{HaTi1986} or for an extension of it, the generalized additive models for location scale and shape (GAMLSS) of \cite{RiSt2005}.

In general, a descent algorithm aims at solving the optimization problem:
$$
\min_{\mathbf{x} \in D} f(\mathbf{x})
$$
with steps of the form $\mathbf{x} \leftarrow \mathbf{x} + t\mathbf{d}
$, moving from the current point $\mathbf{x}$ to the next one. The direction $\mathbf{d}$ and the step size $t>0$ should both be selected such that $f$ is decreasing enough at each step. A common choice for $\mathbf{d}$ is the Newton step $-\{\nabla^2 f(\mathbf{x})\}^{-1} \nabla f(\mathbf{x})$ and $t=1$. Although probably the best choice for convex second-differentiable functions, it is impossible to use in more complex situations, and we do not consider it here. The most common alternative choice is the gradient descent with $\mathbf{d}=-\nabla f(\mathbf{x})$, and $t$ being selected using a line search algorithm. A gradient descent can be applied to more general functions than the Newton descent. However it is much slower to converge.

The local scoring algorithm is a descent algorithm where the final solution should have some predefined properties, often smoothness and additivity in covariates. To achieve this, the descent direction $\mathbf{d}$ is projected onto a set with these properties at each step of the algorithm. The updating rule of the current solution $\mathbf{x}$ is of the form:
\begin{equation}
\mathbf{x} \leftarrow \mathbf{x} + t s(\mathbf{d}), \quad \mbox{ or } \quad \mathbf{x} \leftarrow s(\mathbf{x} + t s(\mathbf{d})),
\label{LS:step}
\end{equation}
where $s$ refers to the projection. 
Replacing the gradient descent direction $\mathbf{d}$ with the gradient sampling approximation looks natural here. 

We present the introduction of the use of the gradient sampling algorithm in two cases. The first is an additive quantile regression. It provides an alternative algorithm to that of \cite{PoKo1997}. The second is a smooth peaks-over-threshold (POT) when the smoothness is imposed on the return levels at different levels instead of the canonical parameters of the generalized Pareto distribution underlying the method. It provides an alternative to GAMLSS, bringing an appreciable stabilization to the maximum likelihood algorithm in that complex setting. 

The remainder of the paper is organized as follows. In Section \ref{GSDA}, we introduce the gradient sampling descent algorithm in general, and combined with the local scoring algorithm for quantile additive models and for POT methods. Two real data cases one in retail store management and the other in the environment context are presented in Section \ref{RealDataExample}. Section \ref{discussion} consists of a discussion. Algorithms are described in details in the Appendix.

\section{The gradient sampling descent algorithm (GSDA)}
\label{GSDA}
\subsection{The algorithm}
The gradient sampling algorithm was first introduced by \cite{BuLeOv2002} and has been further developed by \cite{BuLeOv2005}, \cite{Ki2007}, and \cite{Ki2010}, for example. An up-to-date presentation can be found in \cite{BaKaMa2014}, chap.~13. The idea is to build a gradient descent algorithm where the descent direction is the Clarke subdifferiential. However, the latter cannot be computed exactly because one has to determine the subgradient set, which is feasible only in specific cases. \cite{BuLeOv2002} have therefore proposed the gradient sampling algorithm to approximate it. 

For the sake of completeness, below we quote some definitions following \cite{Ki2007}, adapted to our statistical context. Let $f:\R^n\rightarrow \R$ be a locally Lipschitz continuous function, continuously differentiable\footnote{In the Lebesgue sense, that is the set of point where $f$ is non-continuously differentiable is of null Lebesgue measure.} on an open dense set $D\in \R^n$. The Clarke subdifferential of $f$ at $\mathbf{x}\in \R^n$ is the set:
$$
\bar{\partial}f(\mathbf{x}) = \mbox{conv}\left\{\lim_j \nabla f(\mathbf{z}^j):\mathbf{z}^j\rightarrow \mathbf{x}, \; \mathbf{z}^j \in D\right\},
$$
where conv$(A)$ is the convex hull of $A$. A point $\mathbf{x}$ is called stationary for $f$ if $\mathbf{0} \in \bar{\partial}f(\mathbf{x})$. In particular, when $f$ is $C^1$, then $\bar{\partial}f(\mathbf{x})$ reduces to $\{\nabla f(\mathbf{x})\}$, and a stationary point satisfies $\nabla f(\mathbf{x})~=~0$. The Clarke $\varepsilon$-subdifferential is defined by:
$$
\bar{\partial}_\varepsilon f(\mathbf{x}) = \mbox{conv}\left[\bar{\partial}f\{B(\mathbf{x},\varepsilon)\}\right],
$$
where $B(\mathbf{x},\varepsilon)$ is a ball of radius $\varepsilon \geq 0$ centered at $\mathbf{x}$. A point $\mathbf{x}$ is called $\varepsilon$-stationary for $f$ if $\mathbf{0} \in \bar{\partial}_\varepsilon f(\mathbf{x})$. For a closed convex set $G$, let $\mbox{Proj}(\mathbf{0}|G)$ be the minimum norm element of $G$, or, equivalently, the projection of $\mathbf{0}$ onto $G$. The Clarke generalized subgradient of $f$ at $\mathbf{x}$ is defined by $g_\varepsilon(\mathbf{x})=\mbox{Proj}\{\mathbf{0}|\bar{\partial}_\varepsilon f(\mathbf{x})\}$.  

Because $g_\varepsilon(\mathbf{x})$ cannot be explicitly computed in general, the gradient sampling algorithm approximates it. Denoted by cl conv$(A)$, the closed convex hull of $A$, the Clarke $\varepsilon$-subdifferential $\bar{\partial}_\varepsilon f(\mathbf{x})$ is approximated by:
$$
G_\varepsilon (\mathbf{x}) = \mbox{cl conv}\left[\nabla f\{B(\mathbf{x},\varepsilon)\cap D\}\right],
$$
since $G_\varepsilon(\mathbf{x})\subset \bar{\partial}_\varepsilon f(\mathbf{x})$, and $\bar{\partial}_{\varepsilon_1} f(\mathbf{x})\subset G_{\varepsilon_2}(\mathbf{x})$ for any $0\leq \varepsilon_1 < \varepsilon_2$. To put that in action, the gradient sampling simulates an independent sample $\{\mathbf{x}_i\}_{i=1}^m$ in $\B(\mathbf{x},\varepsilon)$ to approximate $G_\varepsilon(\mathbf{x})$ by 
$$
\hat{G}_\varepsilon(\mathbf{x}) = \mbox{conv} \{\nabla f(\mathbf{x}),\nabla f(\mathbf{x}_1),\ldots,\nabla f(\mathbf{x}_m)\},
$$
where $m \geq n+1$. Next, $g_\varepsilon(\mathbf{x})$ is approximated by $\Hat{g}_\varepsilon(\mathbf{x})=\mbox{Proj}\{\mathbf{0}|\Hat{G}_{\varepsilon}(\mathbf{x})\}$. Finally, the gradient descent algorithm uses the opposite of the Clarke generalized subgradient, $-\hat{g}_\varepsilon(\mathbf{x})$, as the descent direction. 

Henceforth, we will refer to the Gradient Sampling Descent Algorithm using the initialism GSDA. The full algorithm is reported in Appendix~\ref{Appendix}. The main step is the computation of $\hat{g}_\varepsilon(\mathbf{x})$, as follows
\begin{enumerate}
\item Sample $\mathbf{u}_1, \ldots, \mathbf{u}_m$ on the unit ball ${\cal B}(\mathbf{0},1)$, where $\mathbf{0}$ is the $n$-vector of 0.
\item Set $\hat{G}_\varepsilon(\mathbf{x}) = \{\nabla f(\mathbf{x}), \nabla f(\mathbf{x} + \varepsilon \mathbf{u}_1), \ldots, \nabla f(\mathbf{x} + \varepsilon \mathbf{u}_m)\}$.
\item Find 
\begin{equation} \hat{g}_\varepsilon(\mathbf{x}) = \arg \min\left[\Vert g \Vert: g \in \conv \left\{\hat{G}_\varepsilon(\mathbf{x})\right\} \right].
\label{clark}
\end{equation} 
\end{enumerate}
We can see that this step is fairly easy to implement in practice because it only requires to simulate in a ball of small radius around the current $\mathbf{x}$ and to solve a problem that happens to be quadratic and easy to solve. 

\begin{figure}
\centerline{\includegraphics[width=15cm]{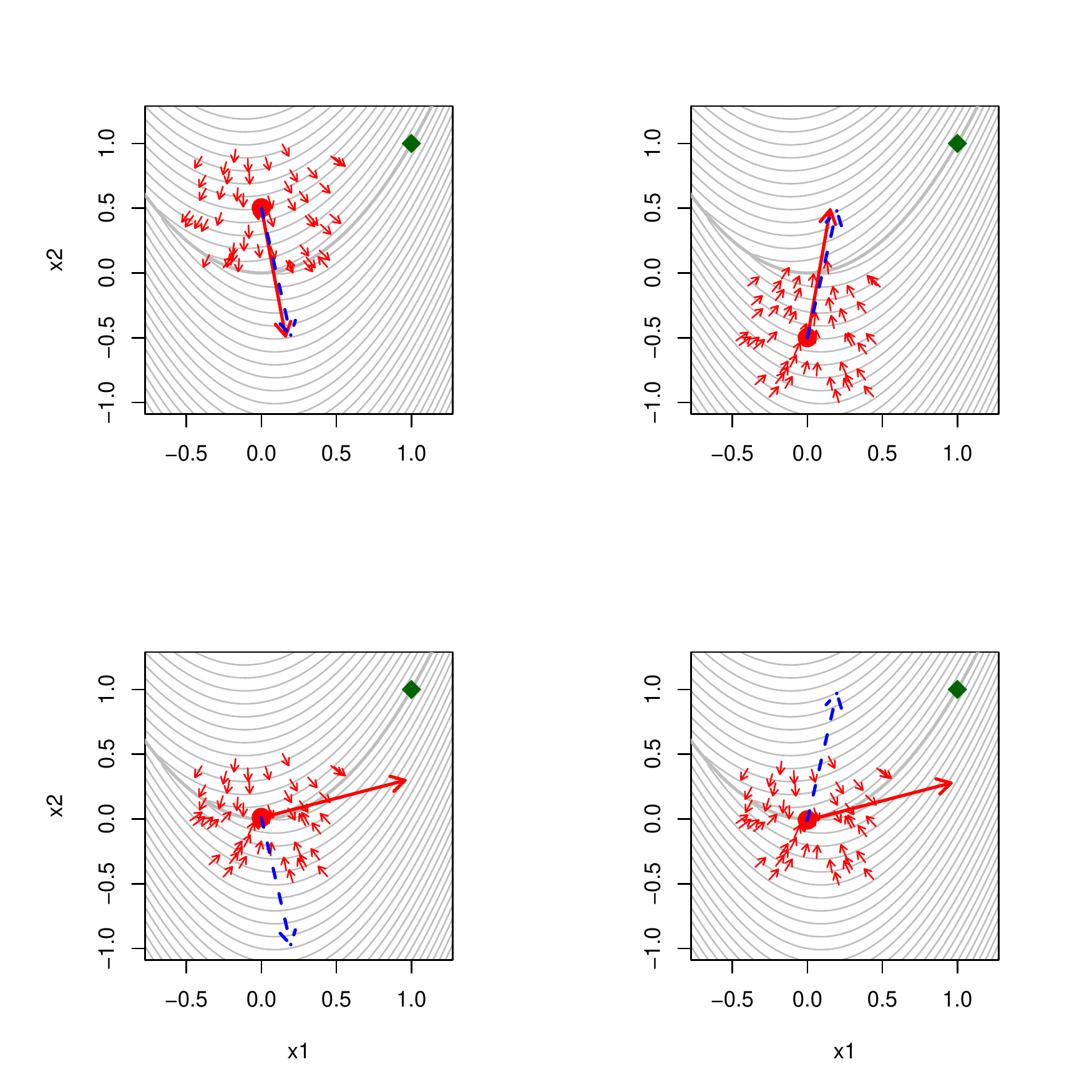}}
\caption{Four cases with function $f(\mathbf{x})=10(x_2-x_1^2)+(1-x_1)^2$. The diamond is at the minimum $(1,1)^T$. The big dot is the current solution $\mathbf{x}$. The bold line is the non-differentiability set $\Omega=\{x_2=x_1^2\}$. The dashed arrow is the gradient descent $-\nabla f(\mathbf{x})$. The short arrows are the sampled gradient descents $-\nabla f(\mathbf{x} + \varepsilon \mathbf{u}_1), \ldots, -\nabla f(\mathbf{x} + \varepsilon \mathbf{u}_m)$. The solid bold arrow is the approximation descent $\hat{g}_\varepsilon(\mathbf{x})$. The lengths of all the arrows have been modified to facilitate readability. Top left plot: current solution $\mathbf{x}$ is far above $\Omega$. Top right plot: $\mathbf{x}$ is far below $\Omega$. Bottom left plot: $\mathbf{x}$ is close above $\Omega$. Bottom right plot: $\mathbf{x}$ is close below $\Omega$.}
\label{GS_illustration}
\end{figure}

Before detailing the GSDA further, let us illustrate how it works at a single step in Figure~\ref{GS_illustration}. Borrowed from \cite{Ov2015}, we consider the function $f(\mathbf{x})=10(x_2-x_1^2)+(1-x_1)^2$. The minimum (diamond) is at $(1,1)^T$. The bold gray line indicates the non-differentiability set $\Omega:=\{x_2=x_1^2\}$. The gradient descent $-\nabla f(\mathbf{x})$ is the dashed arrow. The short arrows are the sampled gradient descents $-\nabla f(\mathbf{x} + \varepsilon \mathbf{u}_1), \ldots, -\nabla f(\mathbf{x} + \varepsilon \mathbf{u}_m)$. The gradient sampling descent $-\hat{g}_\varepsilon(\mathbf{x})$ is the solid bold arrow. The lengths of all the arrows have been modified to facilitate readability. The four plots show cases where the current solution $\mathbf{x}$ (big dot) is either above or below $\Omega$, and either far from it or close to it. In the two top plots, where $\mathbf{x}$ is far from $\Omega$, $\nabla f(\mathbf{x})$ and $\hat{g}_\varepsilon(\mathbf{x})$ are very close: the gradient sampling step is like a descent step when differentiability is guaranteed. In the two bottom plots, where $\mathbf{x}$ is close to $\Omega$, the gradient descent alternates up (right plot) and down (left plot), while the approximation descent is more robust to the current position of $\mathbf{x}$. 

Therefore, in practice, using the gradient sampling at differentiable points close to non-differentiable area on one hand avoids the typical zigzag behavior of descent algorithms. On the other hand, using the gradient sampling comes at the cost of several gradient computations, making it inappropriate in large dimensions or complex cases without further adaptations.

To go into further details, the algorithm requires solving a sub-problem~(\ref{clark}). This can be written as a quadratic problem under linear constraints that can be efficiently solved using classical quadratic programming e.g., function {\tt solveQP} of {\tt R} package {\tt quadprog} \citep{quadprog}. We have noted that, at certain iterations of the algorithm, the sub-problem~(\ref{clark}) may be numerically unstable or difficult to compute. In such case, $\hat{g}_\varepsilon(\mathbf{x})$ can be conveniently replaced with the average of $\{\nabla f(\mathbf{x}), \nabla f(\mathbf{x}+\varepsilon \mathbf{u}_1), \ldots, , \nabla f(\mathbf{x}+\varepsilon \mathbf{u}_m)\}$. Although no formal proof exists, this is intuitive because any stable vector pointing toward $\hat{G}_\varepsilon(\mathbf{x})$ could be used as an approximation of $g_\varepsilon(\mathbf{x})$. The details for solving (\ref{clark}) are reported in Appendix~\ref{Appendix}.

In the two next sections, we show how to apply the association between GSDA and the local scoring algorithm to quantile additive models and to POT methods. In both cases, it consists in using a step as in (\ref{LS:step}), where $\mathbf{d}$ is $-\hat{g}_\varepsilon(\mathbf{x})$.
\subsection{GSDA for quantile additive models}
\label{quantile}
Quantile regression is now a well-known technology that has been introduced in detail, for example in \cite{Ko2005}. An implementation in {\tt R} is available from package {\tt quantreg} with the function {\tt rqss} in the context of additive models for quantile regression \citep{QuReg2016}. Of note, it implements a Frisch--Newton interior point method \citep{PoKo1997}. We now show how using the GSDA provides a simple alternative algorithm. 
 
Let $y_1,\ldots,y_n$ be $n$ observations, independent conditional on $k$-dimensional covariates $\mathbf{w}_1,\ldots,\mathbf{w}_n$, where $\mathbf{w}_i=(w_{i1},\ldots,w_{ik})^T$. In its sample version, the quantile additive model aims at finding $q_\alpha(\mathbf{w}_i) = \alpha_0 + \sum_{j=1}^k f_j(w_{ij})$, a quantile function at level $0< \alpha <1$, minimizing:
\begin{eqnarray*}
\sum_{i=1}^n \rho_\alpha\{q_\alpha(\mathbf{w}_i) ; y_i\},
\end{eqnarray*}
where the risk function is:
$$
\rho_\alpha(q;y) = (1-\alpha)(y-q)^- + \alpha (y-q)^+,
$$   
$z^+ = \max\{z,0\}$, and $z^-=-\min\{z,0\}$. As for any additive model, the functions $f_j$ need to satisfy some identifiability constraints $\sum_i f_j(w_{ij})=0$. At $y\neq q$, the derivative of the risk function is:
\begin{eqnarray}
\nabla \rho_\alpha(q;y) = \left\{
\begin{array}{rl}
1 - \alpha, & \mbox{ if } y-q < 0,\\
- \alpha, & \mbox{ if } y-q > 0.
\end{array}
\right.
\end{eqnarray}
The non-differentiability set of $\rho_\alpha$ makes the classical gradient descent unstable to use in a local scoring. The GSDA offers a conceptually very simple alternative. From the current $\mathbf{q}_\alpha=\{q_\alpha(\mathbf{w}_1),\ldots,q_\alpha(\mathbf{w}_n)\}^T$ , the algorithm moves along $-\hat{g}_\varepsilon(\theta)$ after a smoothing operation. The main steps are as follows:
\begin{enumerate}
\item Sample $\mathbf{u}_1, \ldots, \mathbf{u}_m$ on the unit ball ${\cal B}(\mathbf{0},1)$, where $\mathbf{0}$ is the $n$-vector of 0.
\item Set $\hat{g}_\varepsilon(\mathbf{q}_\alpha) = \left\{\nabla \rho_\alpha(\mathbf{q}_\alpha)+\sum_{k=1}^m \nabla \rho_\alpha(\mathbf{q}_\alpha + \varepsilon \mathbf{u}_k)\right\}/(m+1)$. 
\end{enumerate}
Then
\begin{enumerate}
\item[3.] Set $\mathbf{d}^*= -{\tt gam}(\hat{g}_\varepsilon(\mathbf{q}_\alpha) \sim {\tt lo}(w_1) + \ldots + {\tt lo}(w_k))
$ and $\mathbf{d}=\mathbf{d}^*/\Vert \mathbf{d}^*\Vert$.
\item[4.] Update $\mathbf{q}_\alpha \leftarrow \mathbf{q}_\alpha+t\mathbf{d}$, where $t$ is appropriately selected.
\end{enumerate}
See Appendix~\ref{Appendix} for details, particularly the selection of $t$. The calculation of $\hat{g}_\varepsilon(\mathbf{q}_\alpha)$ can be replaced with the full program (\ref{clark}). The initialization of $\mathbf{q}_\alpha$ can be a constant quantile or a more sophisticated estimate. For the smoothing part $-{\tt gam}(\hat{g}_\varepsilon(\mathbf{q}_\alpha) \sim {\tt lo}(w_1) + \ldots + {\tt lo}(w_k))$, we mimic standard {\tt R} coding and refer the interested reader to \cite{Wo2006} for more details. Note that the smoothing part could be replaced by ${\tt lm}(\hat{g}_\varepsilon(\mathbf{q}_\alpha)\sim w_1+\cdots+w_n)$, for example. This would provide a quantile regression model. 
\subsection{GSDA for peaks-over-threshold method}
\label{GPD}
The POT method, first proposed by \cite{DaSm1990}, is now a standard approach used in extreme value analysis. It has been further developed and refined in many directions. For a review of the methodologies proposed in the non-stationary cases, including additive models, see, for example, \cite{ChDa2005}. 

The POT method assumes that observed excesses, $y_1, \ldots, y_n$,  above a sufficiently high threshold $u$ are independent conditionally on a set of $k$-covariates, $\mathbf{w}_1,\ldots, \mathbf{w}_n$, and are distributed according to a generalized Pareto distribution (GPD) with scale $\sigma_i=\sigma(\mathbf{w}_i)$ and shape $\kappa_i=\kappa(\mathbf{w}_i)$. The log-likelihood function is:
$$
\ell(\sigma, \kappa) = \sum_{i=1}^n -\log \sigma_i - \left(1+1/\kappa_i\right)\log\left(1 + \kappa_i y_i/\sigma_i\right).
$$
The domain of definition is $1+\kappa_i y_i/\sigma_i > 0$, and $\sigma_i > 0$, for $i=1,\ldots,n$. We may assume an additive predictor linked to the parameter such as e.g. \cite{ChDa2005}, 
$$
\sigma_i = h\left\{\beta_0 + \sum_{j=1}^k f_j(w_{ij})\right\}, \quad i=1,\ldots,n,
$$
where $h(\cdot)$ is the link function, $\exp(\cdot)$ in this case, and $f_j$ are smooth functions satisfying identifiability constraints, $\sum_i f_j(w_{ij})=0$. More generally, \cite{RiSt2005} consider any kind of reasonable link functions. In many applications, inference for both the so-called value-at-risk and the expected shortfall is of main importance. For a level $\alpha$, consider:
$$
\theta_i=\theta(\mathbf{w}_i) = \beta^\theta_0 + \sum_{j=1}^p f^{\theta}_j(w_{ij}),\quad
\zeta_i=\zeta(\mathbf{w}_i) = \beta^\zeta_0 + \sum_{j=1}^p f^{\zeta}_j(w_{ij}),
$$
where the value-at-risk $\theta_i\equiv \theta(x_i)$ and the expected shortfall $\zeta_i\equiv \zeta(x_i)$, both at level $1-\alpha$, are respectively:
$$
\theta_i = \left\{
\begin{array}{lr}
(c_\alpha^{-\kappa_i}-1)\sigma_i/\kappa_i,& \kappa_i \neq 0,\\
-\sigma_i \log(c_\alpha),& \kappa_i=0.
\end{array}\right. 
, \quad \zeta_i = \left\{
\begin{array}{lr}
(\theta_i + \sigma_i)/(1-\kappa_i), & \kappa_i\neq 0,\\
\theta_i + \sigma_i, & \kappa_i=0.
\end{array}
\right.
$$
Typically, $1-\alpha=99\%$. The scale factor is $c_\alpha = \alpha/\mbox{Pr}(Y>u)$, and $\zeta_i$ is only defined for $\kappa_i < 1$. 
In the environment context, the value-at-risk is called return level. The interest is in modelling return levels at two different levels, as we illustrate in Section \ref{RealDataExample}, for the U.S. maximal temperatures. 

The GSDA can be used to fit the POT model with value-at-risk additive in the covariates, because it will bring some appreciable numerical stability. For this, the log-likelihood and its derivatives are computed at each step. Denoting $\Theta=(\theta^T,\zeta^T)^T$ and $\Lambda=(\eta^T,\kappa^T)^T$, the derivatives can be conveniently computed with:
\begin{equation}
\nabla_\Theta \ell = \frac{\partial \ell}{\partial \Theta} = \frac{\partial \Lambda^T}{\partial \Theta}\frac{\partial \ell}{\partial \Lambda} = \left(\frac{\partial \Theta^T}{\partial \Lambda}\right)^{-1}\frac{\partial \ell}{\partial \Lambda} = \left(\nabla_\Lambda \Theta\right)^{-1}\nabla_\Lambda \ell.
\label{changevar}
\end{equation}
This provides an explicit formula (i.e., computationally tractable) for the gradient of $\ell$ in $\Theta$. It should be noted at that point that computing $\ell$ and its derivatives given $\Lambda$ is easy, while it is complex given $\Theta$ since one needs to inverse $\Theta(\Lambda)$. 

At each step, $\Theta$ moves along $\mathbf{d}=-s\{\hat{g}_\varepsilon(\Theta)\}$, where $s$ is a suitable transformation for matching the constraint. However, moving in the $\Theta$-space is painful because computing likelihoods and their derivatives is computationally difficult. Instead, we can move in the $\Lambda$-space, using a first-order approximation,
$$
\Lambda^* = \Lambda(\Theta^*) = \Lambda(\Theta + t \mathbf{d}) \approx \Lambda + t \left(\nabla_\Lambda \Theta\right)^{-1} \mathbf{d}.
$$
Overall, the main steps in the algorithm are as follows:
\begin{enumerate}
\item Compute $M=\nabla_\Lambda \Theta$.
\item Compute the approximation $\hat{g}_\varepsilon(\Theta)=\left(\hat{g}_{\varepsilon,\theta}^T(\Theta), \hat{g}_{\varepsilon,\zeta}^T(\Theta)\right)^T$.
\begin{enumerate}
\item Sample $\mathbf{u}_1, \ldots, \mathbf{u}_m$ on the unit ball ${\cal B}(\mathbf{0},1)$, where $\mathbf{0}$ is the $2n$-vector of 0.
\item Set $\hat{g}_\varepsilon(\Theta) = M^{-1}\left\{\nabla_\Lambda \ell(\Lambda)+\sum_{k=1}^m \nabla_\Lambda \ell(\Lambda + \varepsilon \mathbf{u}_k)\right\}/(m+1)$ .
\end{enumerate}
\end{enumerate}
Then,
\begin{enumerate}
\item[3.] Set $\mathbf{d}^*_{\theta}= -{\tt gam}(\hat{g}_{\varepsilon,\theta}(\Theta) \sim {\tt lo}(w_1) + \ldots + {\tt lo}(w_k)).
$
\item[4.] Set $\mathbf{d}^*_{\zeta}= -{\tt gam}(\hat{g}_{\varepsilon,\zeta}(\Theta) \sim {\tt lo}(w_1) + \ldots + {\tt lo}(w_k)).
$
\item[5.] Set $\mathbf{d}^*=((\mathbf{d}^{*}_{\theta})^T,(\mathbf{d}^{*}_{\zeta})^T)^T$, $\mathbf{d}=\mathbf{d}^*/\Vert \mathbf{d}^*\Vert$.
\item[6.] Update $\Lambda \leftarrow \Lambda + tM^{-1} \mathbf{d}$, where $t$ is appropriately selected.
\end{enumerate}
The complete algorithm is reported in Appendix~\ref{Appendix}, in particular some justification for step~2(b), which is an approximation. 
\section{Real data}
\label{RealDataExample}
In this section, we illustrate the use of the gradient sampling algorithm in practice. The first example illustrates the use of the gradient sampling algorithm to estimate a semi-parametric quantile for alimentary products of an European retail store. 
The second shows the estimation of semi-parametric POT return levels at two different levels in the context of temperature data. 
\subsection{European retail store}
\begin{figure}
\centerline{\includegraphics[width=15cm]{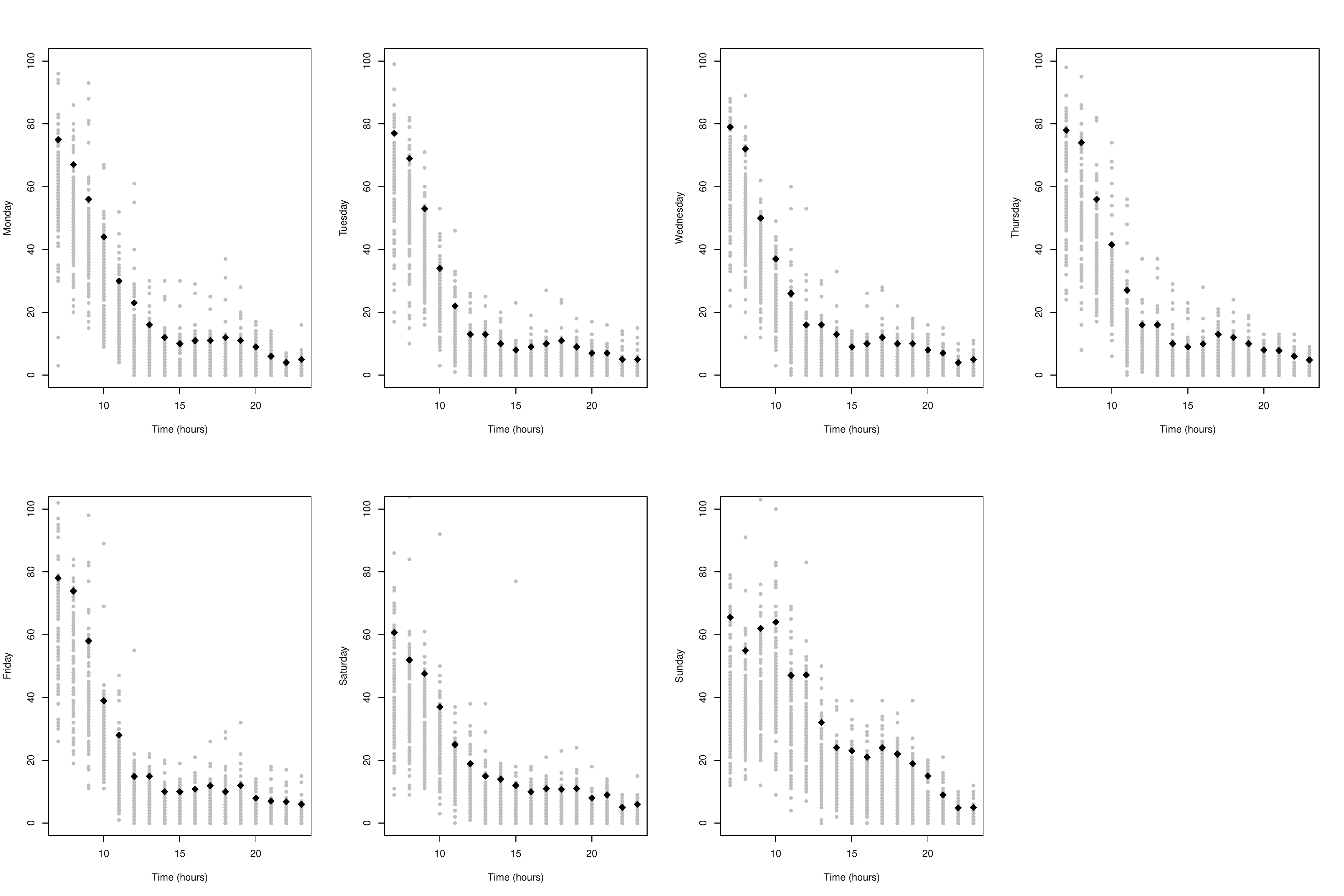}}
\caption{Intradaily sales (points in grey) of butter croissants at the seven days of the week (panels) from November 1, 2012 to November 23, 2014. The black points are the 90\%-quantile estimates.}
\label{ButterCroissant}
\end{figure}

\begin{figure}
\centerline{\includegraphics[width=15cm]{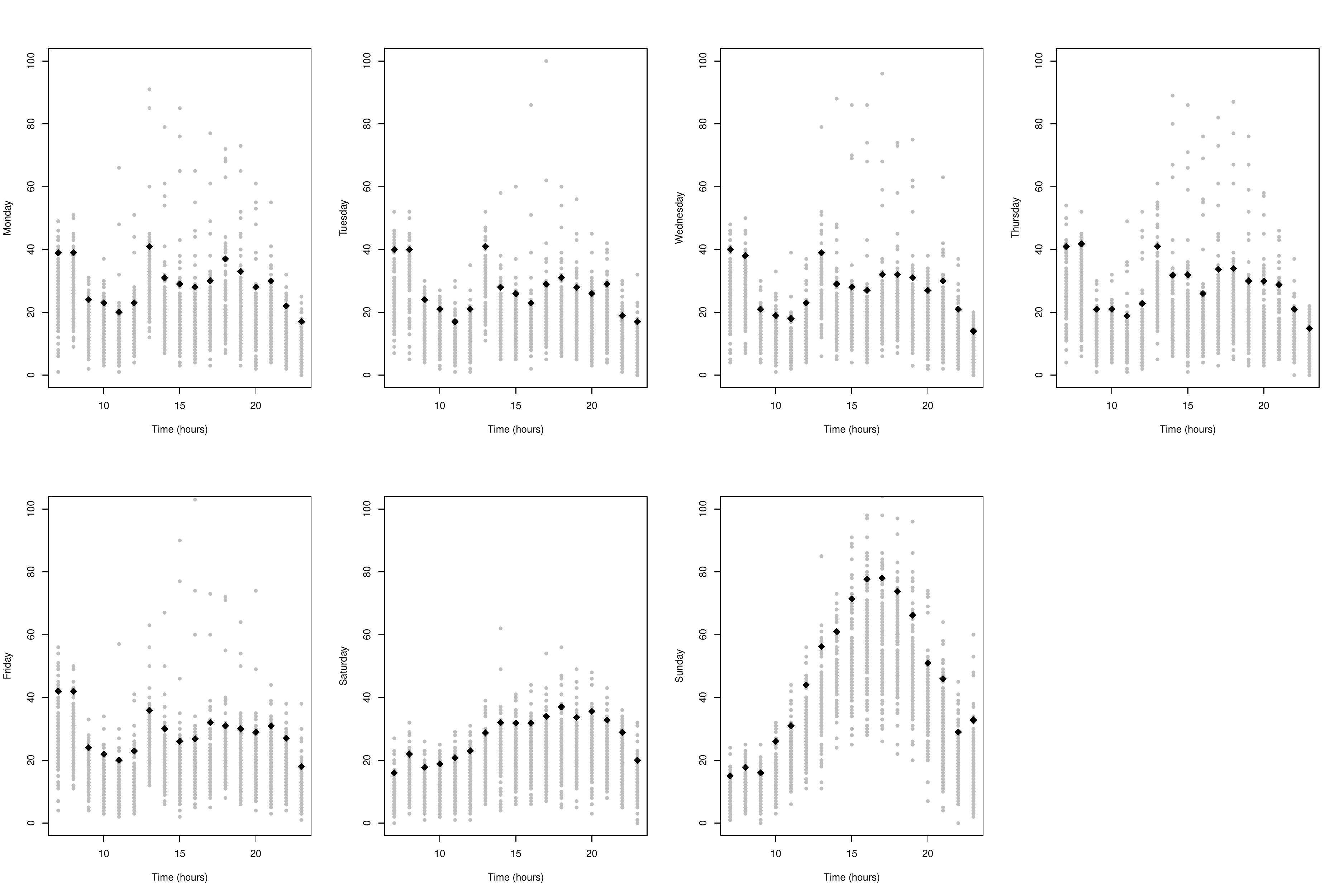}}
\caption{Intradaily sales (points in grey) of energy drink at the seven days of the week (panels) from November 1, 2012 to November 23, 2014. The black points are the 90\%-quantile estimates.}
\label{Energy}
\end{figure}

\begin{figure}
\centerline{\includegraphics[width=15cm]{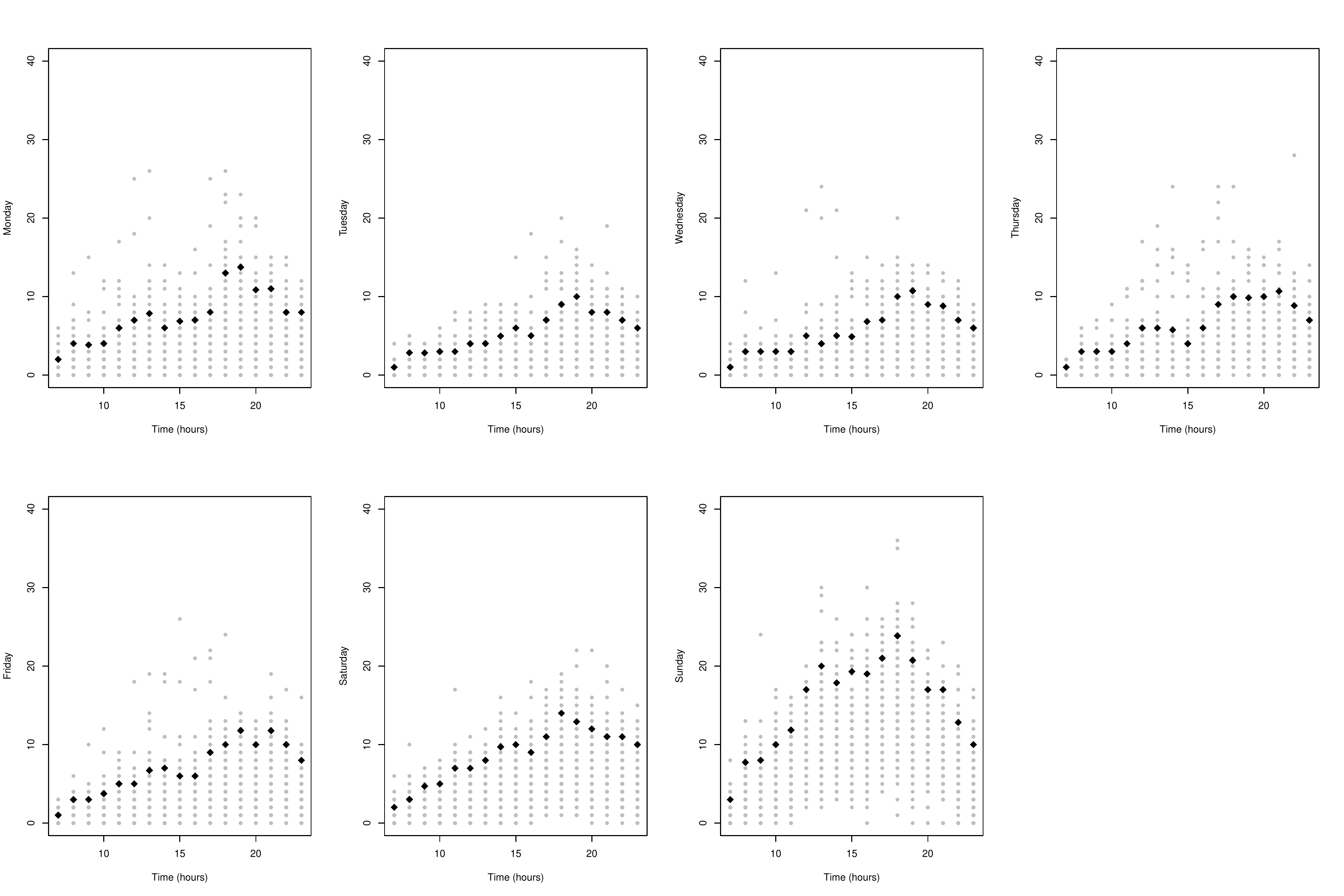}}
\caption{Intradaily sales (points in grey) of milk (one liter) at the seven days of the week (panels) from November 1, 2012 to November 23, 2014. The black points are the 90\%-quantile estimates.}
\label{Milk}
\end{figure}

Shelf replenishment is a large area of operations research \citep{KhGO2008}.  Few works study intradaily sales, although in highly frequented stores some shelves of alimentary products need to be replenished several times a day. This is the case for the  European retail store we consider, which is located in the railway station of a big city in Europe. The store is open every day of the week from 6 am to 11 pm. The data consist of hourly sales of three different products from November 1, 2012 to November 23, 2014, that is 743 days times 17 hours. The products are ``Butter croissants'',  an ``Energy drink'', and ``Milk'' (one liter). The respective daily sales over the period are shown in Figures \ref{ButterCroissant}, \ref{Energy}, and \ref{Milk}. The patterns are different from one product to the other, and Sunday has, in all cases, a clear specific pattern. The croissants are sold mainly early in the morning, with a decreasing trend throughout the day, although a small sales peak occurs again around 5 pm. The energy drinks are basically sold constantly throughout day (except on Sunday), with some peaks appearing at breakfast and lunch time. The sales are much more important on Sunday, especially in the afternoon. The milk sales show an increasing trend until early in the evening for all days. If the store wants to guarantee, at a level of 90\%, that the product shelf is never empty at any time of the day, a quantity of interest is the 90\% quantile of sales. For each product, we use the gradient sampling algorithm given in Section \ref{quantile} to estimate a 90\%-quantile additive model of the form:
$$q_{\alpha}(d_i,h_j)=\alpha_0+ \gamma_{ij} d_ih_j,$$ where $\alpha=90\%$, $d_i$ represents the $i^{th}$-day of the week and $h_j$ the $j^{th}$-hour of the day, and $j=6, \ldots,23$.
The algorithm quickly converges and the results are the black points in Figure \ref{ButterCroissant} for the butter croissants,  Figure \ref{Energy} for the energy drink, and  Figure \ref{Milk} for the one-liter containers of milk. The interaction between day of the week and hour of the day allows the 
90\%--quantile estimate to capture the intradaily and intraweek patterns.  
\subsection{U.S. maximal temperature}
\begin{figure}
\centerline{\includegraphics[width=15cm]{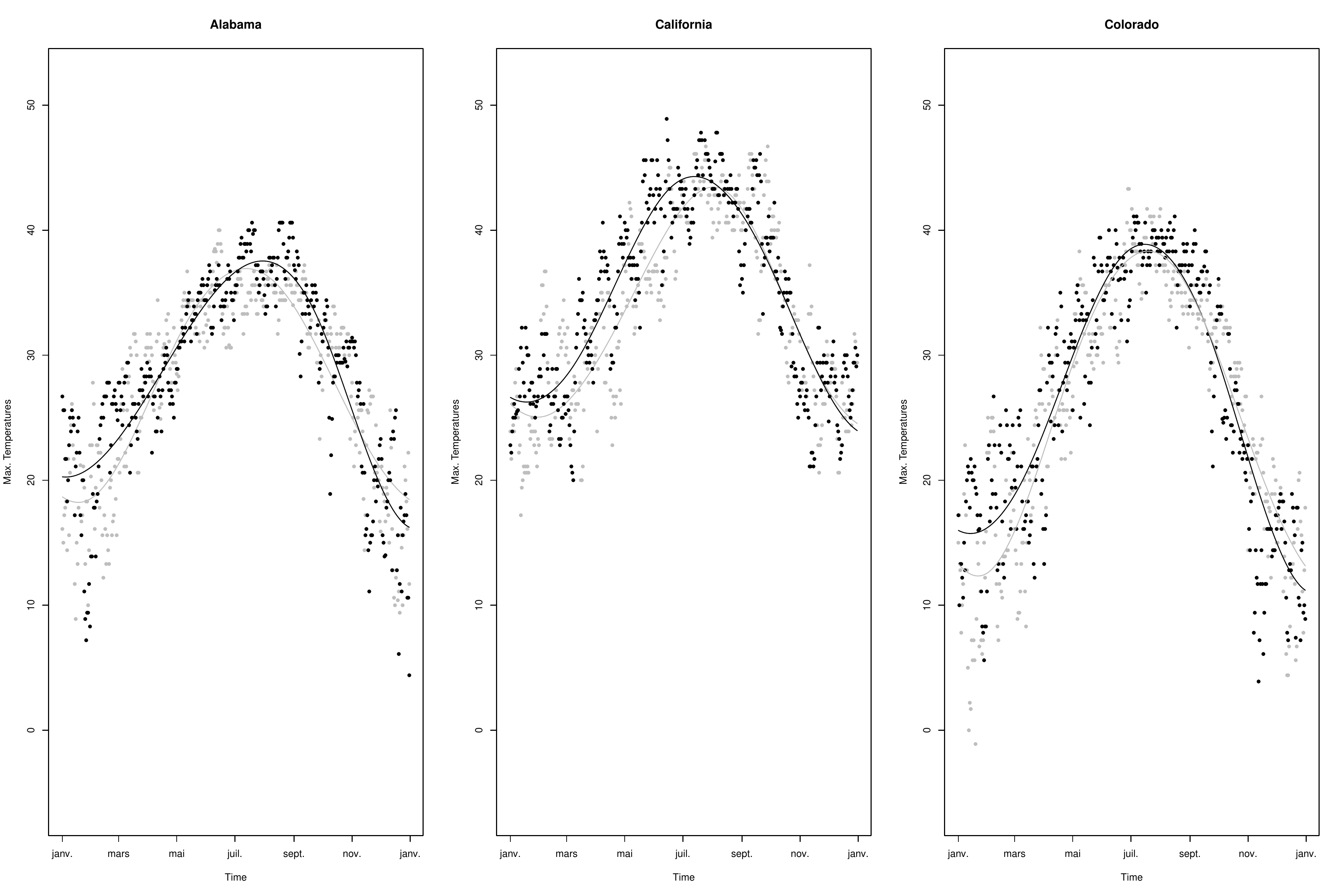}}
\caption{Daily maxima (points) observed at Alabama, California, and Colorado, and the smoothed trend (curve) for 1963 (grey) and 2000 (black).}
\label{MaxTemp}
\end{figure}

\begin{figure}
\centerline{\includegraphics[width=15cm]{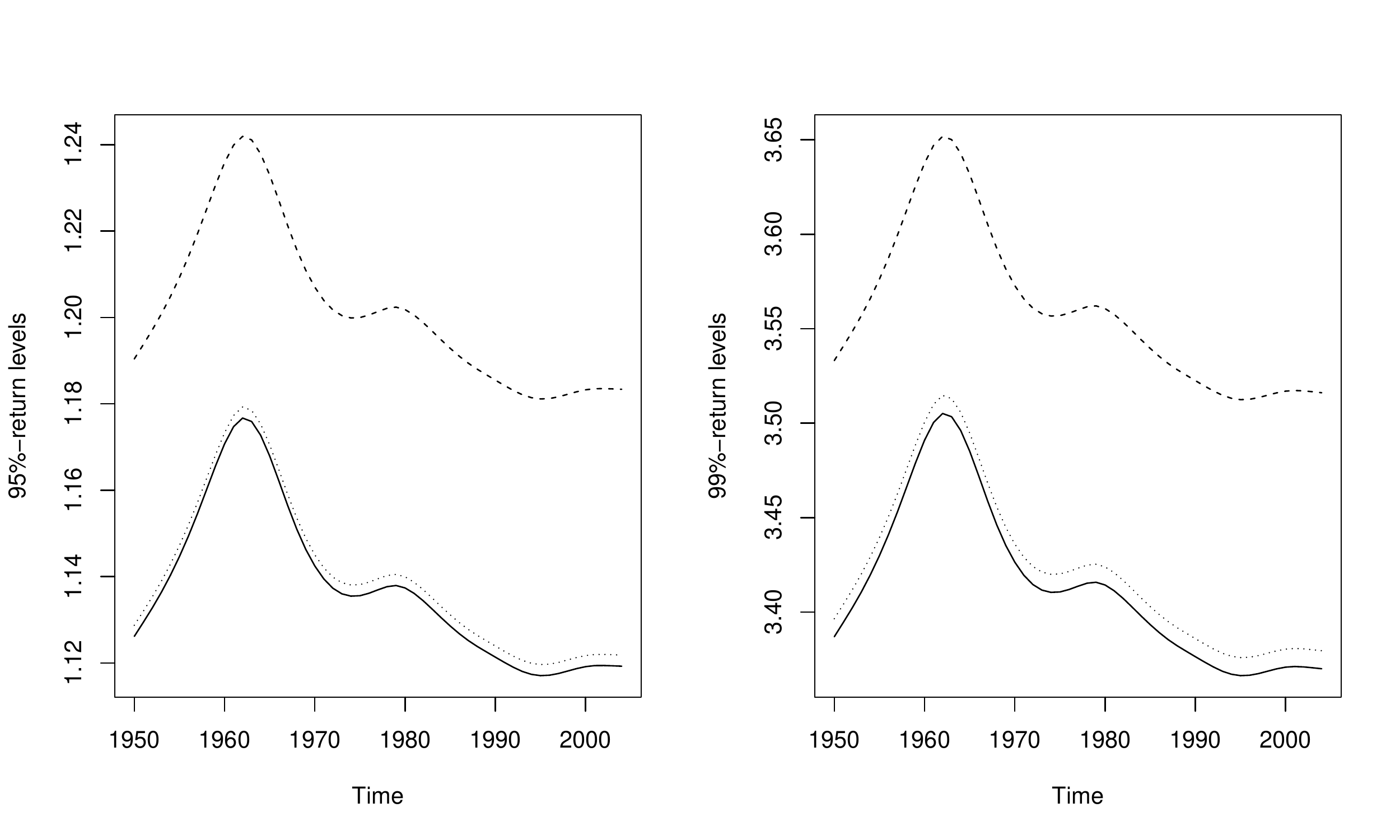}}
\caption{95\% and 99\% return levels estimated at the three sites from 1955 to 2004.}
\label{MaxTempRet}
\end{figure}

We investigate the global change of hot days over the period 1950–2004 at three different stations in the U.S. The data were extracted from the U.S Cooperative Observer Program (COOP) and consist of daily temperature maxima measured  over the period 1950--2004 at different stations in Alabama, California, and Colorado. The dataset can
be freely downloaded from the Climate and Global Dynamics Division (2010). Our aim is to detect the presence of non-stationarity in the return level at two different levels $1-\alpha_1=95\%$, and $1-\alpha_2=99\%$, and its dependence on the station. We detrend the observations of each station by applying a smoothing spline through the 20089 data points and removing the smoothed mean to the observations. Figure \ref{MaxTemp} shows the daily maxima (points) observed at the three sites in 1963 (in grey) and 2000 (black). The line represents the corresponding estimated smoothing trend. The smoothed mean corresponding to year 2000 seems slightly below the one of year 1963. This is later confirmed by the return levels estimates. 
The POT method is the standard quantitative risk methodology when very high quantile calculations are required. Following the POT approach, we fix a constant threshold for each site that corresponds to the 90\%-quantile ot the detrended data and pool together the resulting three site series each of size 2009 over threshold. We fit the GPD return level models as in Section \ref{GPD} with:
$$
\theta_i^{\alpha_1} = \left\{
\begin{array}{lr}
(c_{\alpha_1}^{-\kappa_i}-1)\sigma_i/\kappa_i,& \kappa_i \neq 0,\\
-\sigma_i \log(c_{\alpha_1}),& \kappa_i=0.
\end{array}\right. 
, \quad \theta_i^{\alpha_2} = \left\{
\begin{array}{lr}
(c_{\alpha_2}^{-\kappa_i}-1)\sigma_i/\kappa_i,& \kappa_i \neq 0,\\
-\sigma_i \log(c_{\alpha_2}),& \kappa_i=0.
\end{array}\right. 
$$
 where  $\theta_i^{\alpha_l} = \theta_i^{\alpha_l}(s_i, t_i)$ $l=1,2$ correspond respectively to the $1-\alpha_1=0.95$ and $1-\alpha_2=0.99$ return levels depending on covariates $s_i$ specifying the site and $t_i$ the year of the $i$th value. For simplicity and because of the choice of the threshold, we set $c_{\alpha_l} = \alpha_l/0.01$, although the probability of exceeding the threshold in the denominator could have been estimated by maximum likelihood estimation, which would have led to a value close to 0.01.
 The estimated return levels are shown in Figure \ref{MaxTempRet}. The estimated model uses a GAM with 10 degrees of freedom for the time covariate and an indicator for the site. It shows a global decreasing trend in the sizes of the detrended data above threshold, with higher return levels at  95\% and 99\% for Colorado. The peak of return levels early in the 1960s is in accordance with the findings of \cite{Meetal2009}. The fact that Colorado has a higher return level of size over threshold can be explained by the greater variability of its exceedances over threshold, probably due to altitude. The ability of the methodology to allow a simultaneous estimation of two flexible return levels within the algorithm offers an important clarification in terms of interpretation. First, the two returns levels are likely to behave similarly at different levels. Second, their time-varying structure is directly guided by the data itself and does not resulting from the time-varying forms of the GPD parameters estimated, as would occur in other standard non-stationary models. 
\section{Discussion}
\label{discussion}
In this paper, the gradient sampling algorithm of \cite{BuLeOv2002} is adapted with the local scoring to solve quite complex estimation problems based on the optimization of an objective function (a log-likelihood and a risk function) under some constraints of additivity and smoothness of the parameters. The aim is to overcome non-differentiability and/or non-smoothness of the objective function. In addition, the algorithm can be used for non-convex objective functions at the cost of local convergence.

The algorithm is very flexible and easy to implement. It can be used where a gradient descent could be used if the objective function were differentiable. It can also be used for differentiable functions to reduce numerical instability. The cost is computational because it requires a very large number of calculations of the gradient. This probably precludes its use in several high-dimensional cases. At the time of writing, we are not aware of any improvement to that aspect, although we can be confident that hybrid algorithms will soon bring an efficient solution.

The algorithm was adapted for a quantile additive model, where it offers an alternative algorithm to the Frisch--Newton interior point method of \cite{PoKo1997}, and for the POT model with parameters being smooth and additive in the covariates. Because the final results are maximum likelihood estimators, we have not included any confidence interval calculations, as these are not any different from any that would be obtained with another optimization method.

Even with its drawbacks, the gradient sampling represents a convenient alternative to existing algorithms in many situations where optimization is challenging, and even the only solution in many cases. Therefore, it should be part of the statistician's toolbox.
\section*{Acknowledgements}
This work was supported by the Swiss National Science Foundation.
\bibliography{Methodology}
\appendix
\section{Appendix}
\label{Appendix}
\subsection{Gradient sampling descent algorithm}
\label{GSDA:app}
The gradient sampling algorithm of \cite{Ki2007} is reported below, borrowed from \cite{Ov2015}.
\begin{enumerate}
\item Fix the sampling size $m \geq n+1$, a line search parameter $0 < \beta < 1$, and the reduction factors $\mu\in (0,1)$ and $\lambda \in (0,1)$.
\item Initialize the solution $\mathbf{x}$, the radius $\varepsilon > 0$ and the tolerance $\tau > 0$.
\item Compute the approximation $\hat{g}_\varepsilon(\mathbf{x})$.
\begin{enumerate}
\item Sample $\mathbf{u}_1, \ldots, \mathbf{u}_m$ on the unit ball ${\cal B}(\mathbf{0},1)$, where $\mathbf{0}$ is the $n$-vector of 0.
\item Set $\hat{G}_\varepsilon(\mathbf{x}) = \{\nabla f(\mathbf{x}), \nabla f(\mathbf{x} + \varepsilon \mathbf{u}_1), \ldots, \nabla f(\mathbf{x} + \varepsilon \mathbf{u}_m)\}.$
\item Find 
\begin{equation} \hat{g}_\varepsilon(\mathbf{x}) = \arg \min\left[\Vert g \Vert: g \in \conv \left\{\hat{G}_\varepsilon(\mathbf{x})\right\} \right].
\label{quadprod:app}
\end{equation} 
\end{enumerate}
\item If $\Vert \hat{g}_\varepsilon(\mathbf{x}) \Vert \leq \tau$, update $\varepsilon \leftarrow \mu \varepsilon$ and $\tau \leftarrow \lambda \tau$, go to 3.
\item If $\Vert \hat{g}_\varepsilon(\mathbf{x}) \Vert > \tau$, do a backtracking line search along $\mathbf{d}=-g_\varepsilon(\mathbf{x})/\Vert g_\varepsilon(\mathbf{x}) \Vert$, diminishing $t \in \{1, 1/2, 1/4, \ldots\}$ until the Armijo's condition is satisfied,
\begin{equation}
f(\mathbf{x} + t\mathbf{d}) < f(\mathbf{x})-\beta t \mathbf{d}^T\nabla f(\mathbf{x}).
\label{linesearch}
\end{equation}
\item Update $\mathbf{x} \leftarrow \mathbf{x}+t\mathbf{d}$ and go to 3.
\end{enumerate}
The overall algorithm is stopped when one or several convergence criteria are satisfied, typically, when $\tau$ and $\varepsilon$ reach a predefined threshold.

To solve sub-problem~(\ref{quadprod:app}), recall that any element of $\conv(G)$ can be written as a unique convex combination of vectors at the edges of $\conv(G)$, say $G_e=\{\mathbf{z}_1, \ldots ,\mathbf{z}_m\}$. In more details, let $\mathbf{z} \in \conv(G)$, and then there exists an unique $\mathbf{r}$ such that:
$$
\mathbf{z} = r_1 \mathbf{z}_1 + \dots + r_m\mathbf{z}_l  = Z\mathbf{r}, \quad \mathbf{r} \succeq 0, \quad \sum_{j=1}^m r_j = 1,
$$
where $Z$ is the matrix with columns $\{\mathbf{z}_1, \ldots, \mathbf{z}_l\}$. \cite{Ed1977} shows how to derive the columns of $Z$. The method is efficiently implemented in {\tt chull} of {\tt R}. Because there exits $\mathbf{r}^*$ such that $g=Z\mathbf{r}^*$ for each $g\in \conv \left\{\hat{G}_\varepsilon(\mathbf{x})\right\}$, the sub-problem~(\ref{quadprod:app}) can be written as the quadratic problem under linear constraints:
\begin{eqnarray*}
\min_{\mathbf{r}  \in \R^m} && \mathbf{r} ^T Z^TZ \mathbf{r}\\
s.t. && \mathbf{r} \succeq 0,\\
&& \sum_{j=1}^m r_j = 1.
\end{eqnarray*}
This minimization problem can be efficiently solved using classical quadratic programming (e.g., function {\tt solveQP} of {\tt R} package {\tt quadprog}; \cite{quadprog}). To further simplify this step, we have noted that, at certain iterations of the algorithm, the above problem may be numerically unstable or difficult to compute. In such case, $\hat{g}_\varepsilon(\mathbf{x})$ can be conveniently replaced by the average $(m+1)^{-1} \{\nabla f(\mathbf{x})~+~\sum_{j=1}^m\nabla f(\mathbf{x}+\varepsilon \mathbf{u}_j)\}$. Although we have not found a formal proof, this is intuitive since any stable vector pointing toward $\hat{G}_\varepsilon(\mathbf{x})$ could be used as an approximation of $g_\varepsilon(\mathbf{x})$.

Finally, the Armijo's conditions $(\ref{linesearch})$ can be replaced by $f(\mathbf{x} + t\mathbf{d}) < f(\mathbf{x})-\beta t \Vert \hat{g}_\varepsilon(\mathbf{x}) \Vert$. Indeed, as an approximation  of $\nabla f(\mathbf{x})$, $\hat{g}_\varepsilon(\mathbf{x})$ can be used in the right-hand side of $(\ref{linesearch})$, giving $\mathbf{d}^T\hat{g}_\varepsilon(\mathbf{x}) =  \Vert\hat{g}_\varepsilon(\mathbf{x})\Vert$, see, for instance, \cite{Ov2015}.
\subsection{GSDA for quantile additive models}
The complete algorithm when applied to quantile additive models is given below. 
\begin{enumerate}
\item Fix the sampling size $m \geq n+1$, a line search parameter $0 < \beta < 1$, and the reduction factors $\mu\in (0,1)$ and $\lambda \in (0,1)$.
\item Initialize the radius $\varepsilon > 0$ and the tolerance $\tau > 0$.
\item Initialize the solution $\mathbf{q}_\alpha$.
\item Compute the approximation $\hat{g}_\varepsilon(\mathbf{q}_\alpha)$.
\begin{enumerate}
\item Sample $\mathbf{u}_1, \ldots, \mathbf{u}_m$ on the unit ball ${\cal B}(\mathbf{0},1)$, where $\mathbf{0}$ is the $n$-vector of 0.
\item Set $\hat{g}_\varepsilon(\mathbf{q}_\alpha) = (m+1)^{-1}\left\{\nabla \rho_\alpha(\mathbf{q}_\alpha)+\sum_{k=1}^m \nabla \rho_\alpha(\mathbf{q}_\alpha + \varepsilon \mathbf{u}_k)\right\}.$ 
\end{enumerate}
\item If $\Vert \hat{g}_\varepsilon(\mathbf{q}_\alpha) \Vert \leq \tau$, update $\varepsilon \leftarrow \mu \varepsilon$ and $\tau \leftarrow \lambda \tau$, go to 4.
\item If $\Vert \hat{g}_\varepsilon(\mathbf{q}_\alpha) \Vert > \tau$.
\begin{enumerate}
\item Set $\mathbf{d}^*= -{\tt gam}(\hat{g}_\varepsilon(\mathbf{q}_\alpha) \sim {\tt lo}(w_1) + \ldots + {\tt lo}(w_k)).
$ and $\mathbf{d}=\mathbf{d}^*/\Vert \mathbf{d}^*\Vert$.
\item Diminish $t \in \{1, 1/2, 1/4, \ldots\}$ until $
\rho_\alpha(\mathbf{q}_\alpha + t\mathbf{d}) < \rho_\alpha(\mathbf{q}_\alpha)-\beta t \mathbf{d}^T\nabla \rho_\alpha(\mathbf{q}_\alpha)$.
\end{enumerate}
\item Update $\mathbf{q}_\alpha \leftarrow \mathbf{q}_\alpha+t\mathbf{d}$ and go to 4.
\end{enumerate}
As in Section~\ref{GSDA:app}, the algorithm can be adapted, for example, replacing the average in step~4(b) with the solution to sub-problem~(\ref{quadprod:app}).
\subsection{GSDA for POT}
The complete algorithm when applied to POT is given below. 
\begin{enumerate}
\item Fix the sampling size $m \geq 2n+1$, a line search parameter $0 < \beta < 1$, and the reduction factors $\mu\in (0,1)$ and $\lambda \in (0,1)$.
\item Initialize the radius $\varepsilon > 0$ and the tolerance $\tau > 0$.
\item Initialize the solution $\Lambda=(\eta^T,\kappa^T)^T$, $\Theta=\Theta(\Lambda)=(\theta^T,\zeta^T)^T$.
\item Compute $M=\nabla_\Lambda \Theta$.
\item Compute the approximation $\hat{g}_\varepsilon(\Theta)=\left(\hat{g}_{\varepsilon,\theta}^T(\Theta), \hat{g}_{\varepsilon,\zeta}^T(\Theta)\right)^T$.
\begin{enumerate}
\item Sample $\mathbf{u}_1, \ldots, \mathbf{u}_m$ on the unit ball ${\cal B}(\mathbf{0},1)$, where $\mathbf{0}$ is the $2n$-vector of 0.
\item Set $\hat{g}_\varepsilon(\Theta) = M^{-1}\left\{\nabla_\Lambda \ell(\Lambda)+\sum_{k=1}^m \nabla_\Lambda \ell(\Lambda + \varepsilon \mathbf{u}_k)\right\}/(m+1).$ 
\end{enumerate}
\item If $\Vert \hat{g}_\varepsilon(\Theta) \Vert \leq \tau$, update $\varepsilon \leftarrow \mu \varepsilon$ and $\tau \leftarrow \lambda \tau$, go to 4.
\item If $\Vert \hat{g}_\varepsilon(\Theta) \Vert > \tau$.
\begin{enumerate}
\item Set $\mathbf{d}^*_{\theta}= -{\tt gam}(\hat{g}_{\varepsilon,\theta}(\Theta) \sim {\tt lo}(x_1) + \ldots + {\tt lo}(x_k)).
$
\item Set $\mathbf{d}^*_{\zeta}= -{\tt gam}(\hat{g}_{\varepsilon,\zeta}(\Theta) \sim {\tt lo}(x_1) + \ldots + {\tt lo}(x_k)).
$
\item Set $\mathbf{d}^*=((\mathbf{d}^{*}_{\theta})^T,(\mathbf{d}^{*}_{\zeta})^T)^T$, $\mathbf{d}=\mathbf{d}^*/\Vert \mathbf{d}^*\Vert$.
\item Diminish $t \in \{1, 1/2, 1/4, \ldots\}$ until $\ell(\Lambda + t M^{-1}\mathbf{d}) < \ell(\Lambda)-\beta t M^{-1}\mathbf{d}^T\nabla_\Lambda \ell (\Lambda)$.
\end{enumerate}
\item Update $\Lambda \leftarrow \Lambda + t M^{-1} \mathbf{d}$, $\Theta=\Theta(\Lambda)$, and go to 4.
\end{enumerate}
Focusing on step~5(b), it should be noted that $\hat{g}_\varepsilon(\Theta)$ should be the average of vectors like $\nabla_\Theta \ell(\Theta^*)$ where $\Theta^*=\Theta + \varepsilon \mathbf{u}$, $\mathbf{u}$. We use the following approximation
\begin{eqnarray*}
\nabla_\Theta \ell(\Theta^*) = \frac{\partial \ell}{\partial \Theta}(\Theta^*) 
= \frac{\partial \Lambda^T}{\partial \Theta}(\Theta^*)\frac{\partial \ell}{\partial \Lambda}\left(\Lambda^*\right)
&\approx& \left\{\frac{\partial \Theta^T}{\partial \Lambda}(\Lambda^*)\right\}^{-1}\frac{\partial \ell}{\partial \Lambda}\left[\Lambda + \varepsilon \left\{\frac{\partial \Theta^T}{\partial \Lambda}(\Lambda)\right\}^{-1}\mathbf{u}\right]\\
&\approx& \left\{\frac{\partial \Theta^T}{\partial \Lambda}(\Lambda)\right\}^{-1}\frac{\partial \ell}{\partial \Lambda}\left[\Lambda + \varepsilon \left\{\frac{\partial \Theta^T}{\partial \Lambda}(\Lambda)\right\}^{-1}\mathbf{u}\right].
\end{eqnarray*}
It means that we admit a common matrix $M$ although it could be computed at each sampled points. In addition, given that the distribution of $\mathbf{u}$ on the unit ball ${\cal B}(\mathbf{0},1)$ is not crucial to the algorithm, a further simplification can be obtained by replacing $\left\{\frac{\partial \Theta^T}{\partial \Lambda}(\Lambda)\right\}^{-1}\mathbf{u}$ by $\mathbf{u}$. Finally, like for the general GSDA, the average in step~5(b) could be replaced by the solution to sub-problem~(\ref{quadprod:app}).  








\end{document}